\documentclass{llncs}
\usepackage[latin1]{inputenc}
\usepackage[T1]{fontenc}
\usepackage{amsmath}
\usepackage[pdftex]{graphicx}
\usepackage{url}

\newcommand{\wabs}[1]{\left|#1\right|}
\newcommand{\wcal}[1]{\mathcal{#1}}
\newcommand{\wfc}[2]{{#1}\!\left(#2\right)}
\newcommand{\wref}[1]{(\ref{#1})}
\newcommand{\wvec}[1]{\mathbf{#1}}

\begin{document}
\title{Moore: Interval Arithmetic in C++20}
\author{W.~F.~Mascarenhas}
\institute{University of S\~{a}o Paulo, Brazil,\\
\email{walter.mascarenhas@gmail.com}}

\markboth{Submitted to the 37th Annual Meeting of the NAFIPS}
{Moore: Interval Arithmetic in modern C++}


\maketitle

\begin{abstract}
This article presents the Moore library for interval arithmetic in C++20. 
It gives examples of how the library can be used, and explains the basic
principles underlying its design. 
\end{abstract}


\section{Introduction}
\label{sec:introduction}

This article presents the Moore library for interval
arithmetic in C++20. It gives examples of how
the library can be used, and explains the basic
principles underlying its design. It also
describes how the library differs from the
several other good libraries already available
\cite{boost,cxsc,filib,gaol,gnu,jinterval,lambov,mpfi,nehmeierA,profil,rump,sun}.
The Moore library is not compliant with the recent IEEE standards
for interval arithmetic \cite{full,p17881},
and it will never be, but it would fair to
rank in the top five in terms of compliance among the
libraries in \cite{boost,cxsc,filib,gaol,gnu,jinterval,lambov,mpfi,nehmeierA,profil,rump,sun},
the first and only truly compliant being \cite{gnu}, followed by \cite{nehmeierA},
which is almost compliant.
Of course, the library has limitations, and some of them
are addressed in the last section, but only by playing a bit
with it you will be able to tell whether
its pluses offset its minuses.

The library was written mainly for myself
and my students, to be used in
our research about interval arithmetic
and scientific computing in general.
It is also meant to be used by other people,
and its open source code and manual are available
upon request to me. It is distributed
under the Mozilla 2.0 license.

The Moore library will be useful
for people looking for better performance or
more precise types of endpoints for their intervals.
To emphasize this point, Section \ref{sec:experiments} presents
experiments showing that it is competitive
with well known libraries, and it is significantly faster
than some of them. The library will be most helpful
for people using single or double precision
arithmetic for most of their computation, with sporadic
use of higher precision to handle critical particular cases.
In this scenario the Moore library offers tools
which are not available ``out of the box'' in
other libraries, if available at all.

I assume that you are familiar with interval arithmetic,
and understands me when I say that the library satisfies all
the containment requirements of interval arithmetic.
I also assume that you have experience with templates,
but you do not need to be familiar with
the feature of C++20 which distinguishes
most the Moore library from the others:
{\it Concepts} \cite{cpts}, which are described in Section
\ref{sec:concepts}.

In the rest of the article I present
the library, starting from the basic
operations and moving to more advanced features,
and present extensions of the library
for linear algebra and automatic differentiation.

\section{Hello Interval World}
\label{sec:hello}
The Moore library can be used by people
with varying degrees of expertise. Non
experts can simply follow what is outlined
in the code below:
{\small
\begin{verbatim}
#include "moore/config/minimal.h"
...
using namespace Ime::Moore;
...
UpRounding r;
TInterval<> x(2.0, 3.0);
TInterval<> y("[-1/3, 2/3]");

for(int i = 0; i < 10; ++i) {
  y = (sin(x) - (y/x + 5.0) * y) * 0.05;
  cout << y << endl;
}
\end{verbatim}
}
\noindent
With the Moore library you construct intervals
by providing their endpoints
as numbers or strings, and then use
them in arithmetical
expressions as if they were numbers. The
library also provides trigonometric and
hyperbolic functions, their inverses,
exponentials and logarithms,
and convenient ways to read and write
intervals to streams.

The file \texttt{/moore/config/minimal.h }
contains the required declarations for
using the library with \texttt{double} endpoints.
The line
{\small \
\begin{verbatim}
UpRounding r;
\end{verbatim}
}
\noindent
is mandatory. It sets the rounding mode to upwards,
and the rounding mode is restored
when \texttt{r} is destroyed. This is like one of the
options in the boost library \cite{boost}, but
the Moore library uses only one rounding policy. In fact,
giving fewer options instead of more is my usual choice.
I only care about concrete use cases motivated by
my own research, instead of all possible uses of interval arithmetic.
I prefer to provide a better library for a few users rather than
trying to please a larger audience which
I will never reach.

Intervals are represented by the
class template \texttt{TInterval<E>}, which is
parameterized by a single type \texttt{E}.
The letter \texttt{E} stands for {\it endpoint},
and both endpoints of the same interval are
of the same type \texttt{E}, but intervals
of different types may have different types of endpoints,
and we can operate with them, as illustrated below.
The default value for \texttt{E} is \texttt{double}, so that
\texttt{TInterval<>} represents the plain vanilla
intervals with endpoints of type \texttt{double}
available in other libraries.

The library does not contain class hierarchies,
virtual methods or policy classes.
On the one hand, you can only choose
the type of the endpoints defining the intervals
of the form $[a,b]$ with $-\infty \leq a \leq b \leq +\infty$,
or the empty interval.
On the other hand, I do believe that it goes
beyond what is offered by other libraries
in its support of generic endpoints, intervals and operations.
The library works with several types of endpoints
``out of the box,'' that is, it
provides tested code in which several
types of endpoints can be combined,
as in the example below.
It also implements other kinds of
convex subsets of the real line. For instance, it
has classes to represent intervals of the form
\texttt{(a,b]}, \texttt{[a,b)} or \texttt{(a,b)},
in which the ``openness'' of the endpoints can
be decided at compile or runtime, and these
half open intervals are used to implement
tight arithmetic operations.

The code below illustrates the use of intervals
with four types of endpoints:
{\small
\begin{verbatim}
TInterval<>           x(5,6);
TInterval<float>      y(-1,2);
TInterval<__float128> z("[-inf,4"]);
TInterval<RealEnd<256>> w("[-1/3,2/3]");

auto h = x | y | 0.3;                // the convex hull of x,y and 3
auto i = x & y & z & w;              // the intersection of x,y,z and w
auto j = sin(z * x/cos(y * z)) - exp(w);
\end{verbatim}
}
\noindent
\begin{itemize}
\item The interval \texttt{x} has endpoints of type \texttt{double}.
\item \texttt{y} has endpoints of  type \texttt{float}.
\item The endpoints of \texttt{z} have quadruple precision.
\item \texttt{w} has endpoints of type \texttt{RealEnd<256>}, which
are floating point numbers with $N=256$ bits of mantissa,
and you can choose other values for $N$.
\item The compiler deduces that \texttt{h} is an interval with endpoints of type
\texttt{double}, which is the appropriate type for storing the convex hull of
\texttt{x}, \texttt{y} and \texttt{0.3}.
\item It also deduces that \texttt{RealEnd<256>} is the appropriate type of endpoints
for the interval representing the intersection of \texttt{x}, \texttt{y}, \texttt{z} and
\texttt{w}, and this is the endpoint type for \texttt{j}.
\end{itemize}

I ask you not to underestimate the code above. It
is difficult to develop the infrastructure required  to
handle intervals with endpoints of
different types in expressions as natural
as the ones in that code.
In fact, there are numerous
issues involved in dealing with intervals
with generic endpoints, and simply writing
generic code with this purpose is not enough.
The code must be tested, and my experience
shows that it may compile for some types
of endpoints and may not compile for others.

\section{Concepts}
\label{sec:concepts}
The Moore Library differs significantly
from the previous C++ interval arithmetic libraries
due to its use of {\it Concepts},
a feature which will be part of the C++20 standard \cite{cpts}.
Concepts improve the diagnostic of errors in the compilation of templated
C++ code, and they can be motivated by the following example.
Suppose we write the code below to compute the length
of intervals of types \texttt{Interval} provided by several
libraries.
{\small
\begin{verbatim}
template <typename Interval>        // Code in a header file somewhere.
double length(Interval const& i) {  // Interval is meant to be a type
  return sup(i) - inf(i);           // provided by an existing interval
}                                   // arithmetic library.
\end{verbatim}
}
\noindent
This code works as long as the functions \texttt{inf} and \texttt{sup}
are provided, either by the original library for the
type \texttt{Interval} or by an adapter.
However, it will not take long for someone
to code something like the snippet below and get
indecipherable error messages about
\texttt{inf}s, \texttt{sup}s and strings.
{\small
\begin{verbatim}
void unlucky()      // code in a source file unrelated to intervals.
{
  std::string str("I know nothing about intervals!!!");
  std::cout << length(str) << std::endl;
}
\end{verbatim}
}
\noindent
When reading the error messages about
\texttt{inf}s and \texttt{sup}s of strings in the
compilation of the \texttt{unlucky} function, the programmer may
not be aware of the chain of inclusions
leading to the header file containing
the declaration of the function \texttt{length} for
intervals, and the \texttt{length} function for strings will
be declared in yet another header file.
It will be difficult to
relate the error messages to the code which is
apparently being compiled, and unexperienced programmers
will get lost.  Even people experienced with
templates will tell you how
frustrating these error messages can be, and this
is indeed a problem with templates.

We could solve this problem by telling
the compiler what an interval is.
Knowing that strings are not intervals,
it would not consider
the function template \texttt{length} below as an option for strings,
and there would be no meaningless error messages
about \texttt{inf}s and \texttt{sup}s of strings.
{\small
\begin{verbatim}
template <Interval I>         // Telling the compiler that I must be an
double length(I const& i) {   // interval for this function template to
  return sup(i) - inf(i);     // be considered.
}
\end{verbatim}
}
In essence, this is what a concept is: a way to
tell the compiler whether a type should be
be considered in the instantiation of templates.
In the Moore library
concepts are used, for example, to tell whether a type
represents an interval (the \texttt{Interval} concept)
or an endpoint (the \texttt{End} concept), or when
there exist an exact conversion from endpoints
of type \texttt{T} to endpoints of type E
(the \texttt{Exact<T,E>} concept.)
We then can code as follows and the compiler will
instantiate the appropriate templates. In the
end, concepts allow us to operate naturally
with intervals and endpoints of different types.
{\small
\begin{verbatim}
template <Interval I>            // sum of intervals of the same type
I operator+(I const&, I const&)
\end{verbatim}
}
{\small
\begin{verbatim}
template <Interval I, Interval J>     // sum of intervals when there
requires Exact<EndOf<J>, EndOf<I>>()  // is an exact conversion from
I operator+(I const&, J const&)       // J to I.
\end{verbatim}
}
{\small
\begin{verbatim}
template <Interval I, Interval J>     // sum of intervals when there
requires Exact<EndOf<I>, EndOf<J>>()  // is an exact conversion from
J operator+(I const&, J const&  )     // I to J.
\end{verbatim}
}
{\small
\begin{verbatim}
template <Interval I, End E>      // sum of an intervals and an
requires Exact<E, EndOf<I> >()    // endpoint when there is an
I operator+(I const&, E const&)   // exact conversion form E to I.
\end{verbatim}
}
\noindent
The code above also presents an alternative way
to enforce concepts: the \texttt{requires} clauses. These
clauses make sure that the \texttt{operator+}
will be considered only when there is an obviously
consistent type for the output.

Overall, the motivation for concepts is clear and intuitive.
Their problems lie in the details and the
crucial question: How should we tell the compiler what
an interval or and endpoint is (or any concept, really)?
I do not know the best answer to this question,
and neither does the rest of the C++ community.
This is why concepts are taking
so long to become part of the C++ standard.

This ignorance should not prevent us from using concepts.
They are a great tool, and we can do a lot with what
is already available. With time, as
concepts and our experience with them evolve, we will
improve our code. For now the Moore library tells the compiler
in an ad-hoc way what intervals and endpoints
are. It basically lists explicitly which types
qualify for a concept, and avoids the more elaborate schemes
to declare concepts which are already available,
for two reasons: First, their current implementation
has bugs (it does not handle recursion properly,
for instance.) Second, it is difficult to list precisely
and concisely all the requirements which would
characterize intervals and endpoints. I would not
be able to do it even if the current implementation
of concepts were perfect.

The last questions are then: do concepts work
for interval arithmetic? Are they worth the trouble?
I would not have written this article if my answer to
these questions were not an enthusiastic ``yes!!'',
and I invite you to try out the library
and verify whether you share my enthusiasm.

\section{Input and output}
\label{sec:io}
Flexible and precise input and output are essential
for an interval arithmetic libary. The Moore library accepts
as input interval literals and streams as follows
{\small
\begin{verbatim}
try {
  TInterval<> x("[]");       // the empty interval
  x = "[-inf, 1]";           // -inf = minus infinity
  x = "[2.0e-20, 1/3]";      // rational numbers are ok
  x = "[-2.345, 0x23Ap+4]";  // hexadecimal floats too
  std::cin >> x;             // reading from an input stream
} catch(...){}
\end{verbatim}
}
\noindent
As the code above indicates, the library throws
an exception when the string literal meant to
represent an interval is invalid.
Strings in hexadecimal notation are handled exactly,
and by using them for both input and output you can
persist intervals without rounding errors. In the
other formats the resulting interval is usually the tightest
representable interval containing the input, the only
exception being contrived rational numbers for which it
would take an enormous amount of memory or time to
compute this tight enclosure. In these rare cases
you may get a memory allocation exception or
need to wait forever.

Properly formatted output is important to visualize
the results of interval computations, and the
library implements an extension of the usual
printf format to specify how intervals are
written to streams. This extension is
needed in order to align numbers properly in
columns when printing vectors and matrices.
For example, the code below creates a $3 \times 3$
matrix of intervals (a box matrix) and writes it
to the standard output. The output is formatted
according to the string \texttt{"11.2E3W26"}, which
extends the argument
\texttt{"+11.2E"} passed to
\texttt{printf}  to write floating point
numbers in scientific notation (E),
showing the plus sign (+), with 11 characters
per number and 2 digits after the decimal point.
We add the suffix "3W26" to the format to
ensure that exponents are printed with 3 digits
and each interval is 26 characters wide. Without this
extension the output would not be as well
as organized at it is below.
{\small
\begin{verbatim}
using I = TInterval<>;
text_format() = "+11.2E3W26";
TBoxMatrix<> a( { { I(0x1p-1021,0x1p+100), I()          },
                  { I("[-inf,0]"),         I("[0,inf]") },
                  { I(-12343,0),           I(50,10000)  } } );
std::cout << a << std::endl;
\end{verbatim}
}
\noindent
This is the output:
{\small
\begin{verbatim}
 [ +4.45E-308, +1.27E+030] [                       ]
 [       -INF, +0.00E+000] [ +0.00E+000,       +INF]
 [ -1.24E+004, +0.00E+000] [ +5.00E+001, +1.00E+004]
\end{verbatim}
}

\section{Linear Algebra}
\label{gibbs}
Besides plain intervals, the library provides
vectors of intervals, called boxes, and matrices with
interval entries (box matrices) The arithmetic operations
involving vectors and matrices are implemented
using expression templates and one can write code as the
one below,  which handles the three dimensional vectors
\texttt{x} and \texttt{y} and the $3 \times 3$ matrix
\texttt{a} in a natural way.
{\small
\begin{verbatim}
using I = TInterval<>;

TBox<> x( {I(1,3), I(2,4), I(1,5)} );
TBox<> y( {I(1,2), I(2,3), I(2,3)} );

TBoxMatrix<> a( { { I(1,1), I(0,1), I(3,5) },
                  { I(2,1), I(2,2), I(4,7) },
                  { I(2,1), I(2,2), I(3,5) } });

TBox<> z = a * x + 2 * y + x;
TBox<> w = tr(a) * y + dot(y,z) * x; // tr(a) = transposed(a)

\end{verbatim}
}

\section{Automatic Differentiation}
\label{taylor}
The Moore library is part of a larger collection
of tools for scientific computing, called
{\it Ime library}. As part of the work of my student Fernando Medeiros,
the Ime library provides classes for automatic
differentiation, and I now describe how these automatic differentiation tools
by Fernando and myself are integrated with the Moore library.
First, we use a function template  to declare
the function which we want to differentiate.
{\small
\begin{verbatim}
template <typename T>
T example(T const& x) {
  return exp( sqrt(exp(x)/ 3) + x) * (2 * x) - 10;
}
\end{verbatim}
}
\noindent
Once we have declared \texttt{example}, it is easy to compute
its derivative using interval arguments.
For instance, the function \texttt{newton\_step}
below performs one step of Newton's method
for solving the equation $\wfc{f}{x} = 0$.
In this code the type \texttt{ADT<I>} represents
the usual pair of function value and derivative
used in forward automatic differentiation schemes.
{\small
\begin{verbatim}
template <Interval I>
void newton_step(I& x, ADT<I> (*f)(I const& i)) {
  auto fd = adt(x, example); // evaluating f and its derivative
  x &= x - fd.f / fd.d;      // x = (x - f(x)/f'(x)) intersected with x
}

void calling_newton() {
  TInterval<> x(1,2);
  newton_step(x, example);
}
\end{verbatim}
}

The library Ime also provides automatic differentiation for functions
of several variables, like in the example
below in which we print the enclosure of
the function value and gradient of the given
multivariate function.
{\small
\begin{verbatim}
template <typename T, int N>
T multivariate_example(StaticVector<T,N> const& x) {
   return exp( sqrt(exp(x[0] + x[1] / 3) + x[2]) * (2 * x[3])) / x[4];
}

void print_function_value_and_gradient() {
  using I = TInterval<>;
  text_format() = "+10.4E";
  StaticVector<I,5> x( {I(1,2), I(-2,3), I(3,4), I(-1,1), I(1,2)} );
  std::cout << adtnf(x, multivariate_example);
}
\end{verbatim}
}
\noindent
This is the output:
{\small
\begin{verbatim}
   f = [+2.730E-05,+1.832E+04]
g[0] = [-3.509E+05,+3.509E+05]
g[1] = [-1.170E+05,+1.170E+05]
g[2] = [-1.748E+04,+1.748E+04]
g[3] = [+5.724E-05,+3.596E+05]
g[4] = [-1.832E+04,-1.365E-05]
\end{verbatim}
}
\section{Experiments}
\label{sec:experiments}
This section presents the results of experiments comparing
the Moore library with three other interval arithmetic libraries:
boost interval \cite{boost}, Filib \cite{filib}
and libieeep1788 \cite{nehmeierA}. In summary,
the experiments show that, for arithmetic operations,
the Moore library is slightly faster than the
boost library, it is significantly faster than the libieeep1788 library, and
it is faster than the Filib library. However, in double
precision the elementary functions
(sin, cos, etc.) in Filib are significantly faster than the Moore
library, which is in turn significantly faster than the boost library
and the libieeep1788 library.

\begin{table}[!t]
\renewcommand{\arraystretch}{1.3}
\caption{Normalized Times for the Lebesgue Function}
\label{table:lebesgue}
\centering
\begin{tabular}{cccc}
\hline
\texttt{Moore} & \texttt{Filib} & \texttt{boost} & \texttt{P1788}\\
1     & 3.8   & 1.1   & 268.5 \\
\hline
\end{tabular}
\end{table}

Besides the difference in speed, there is a difference in
the accuracy of the elementary functions. When using
IEEE754 double precision, due to the way
in which argument reduction is performed, the boost
and Filib libraries can lead to errors of
order $10^{-8}$
in situations in which the Moore library and the libieeep1788 library
lead to errors of the order $10^{-16}$.
In fact, in extreme cases these other
libraries can produce intervals of length 2 when
the sharpest answer would be an interval of length
of order $10^{-16}$.

The Moore library was implemented to be used
in my research, and the experiments reflect this. I present timings
related to my current research about
the stability of barycentric interpolation \cite{MascA,MascB,MascC}.
In this research I look for parameters $w_0,\dots w_n$ which
minimize the maximum of the {\it Lebesgue function}
\begin{equation}
\label{lebesgue}
\wfc{\wcal{L}}{\wvec{w};\wvec{x},t} :=
\ \ \left.
\sum\limits_{k = 0}^{n} \wabs{\frac{w_k}{t - x_k}} \right/ \wabs{\sum\limits_{k = 0}^{n} \frac{w_k}{t - x_k}}
\end{equation}
among all $t \in [-1,1]$, for a given vector $\wvec{x}$ of nodes,
and I use interval arithmetic to find such minimizers and
validate them.

The first experiment timed the evaluation of the
Lebesgue function for 257 Chebyshev nodes
of the second kind \cite{MascA}, with interval weights,
at a million points $t$. I
obtained the normalized times in Table \ref{table:lebesgue} (the time for
the Moore library was taken as the unit.) This table indicates that
for the arithmetic operations involved in the evaluation of the
Lebesgue function \wref{lebesgue} the Moore library is more
efficient that the boost, Filib and libieeep1788 libraries.
The difference is slight between Moore and boost (10\%),
more relevant between Moore and Filib (about 300\%) and
very significant between Moore and libieeeP1788 (about 25000\%).

\begin{figure}[!t]
\centering
\includegraphics[height=6.0cm, width=9.0cm]{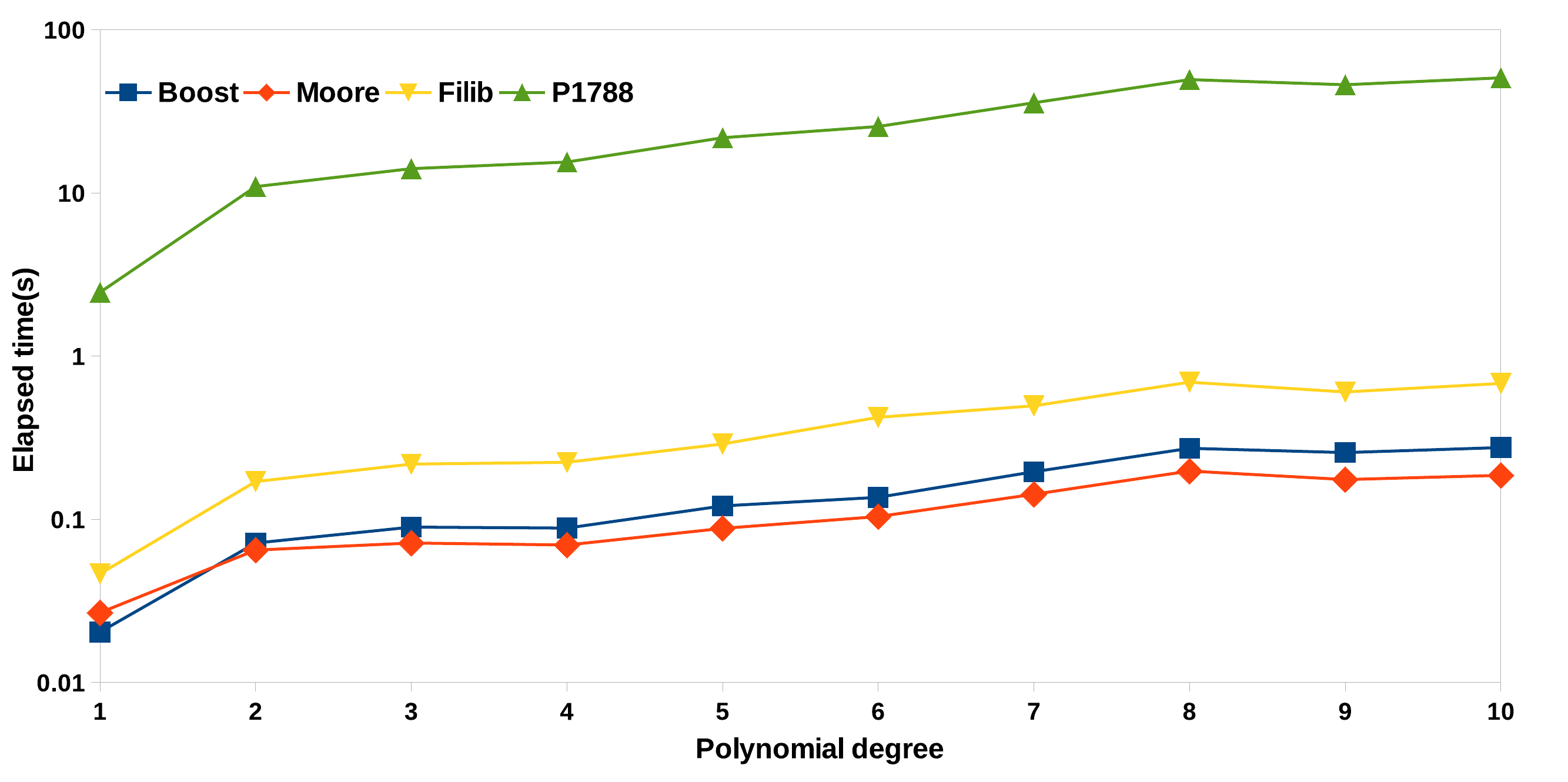}
\caption{Times for Newton's method with polynomials, in log scale.}
\label{fig:newton}
\end{figure}

In the second experiment, myself and my former
student Tiago Montanher considered the computation of the roots of
functions which use only arithmetic operations, like the
Lebesgue function in Equation \wref{lebesgue} and its derivatives
with respect to its parameters.
The data for this experiment was generated with
an interval implementation of Newton's method
which can use any one of the four libraries mentioned above.
We compared the times for the solution of random polynomial
equations, with the polynomials and their derivatives
evaluated by Horner's method. We obtained the times
in Figure \ref{fig:newton}, which corroborate the
data in Table \ref{table:lebesgue}.

The first two experiments show that the Moore library is competitive
for arithmetic operations, but they tell only part of the history about
the relative efficiency of the four libraries considered. In order to
have a more balanced comparison, in the third and last experiment
I compared the times that the four libraries mentioned above take to
evaluate of the elementary functions (sin, cos, exp, etc.)
using the IEEE 754 double precision arithmetic.
The results of this experiment are summarized in Table \ref{table:elem} below,
which shows that the Filib library is much
faster than the Moore library in this scenario, and the
Moore library is much faster than the other two libraries.

\begin{table}[!t]
\caption{Time for $10^6$ evaluations of the elementary functions with random intervals.}
\label{table:elem}
\centering
\setlength{\tabcolsep}{0.45em}
\begin{tabular}{|l|r|r|r|r|}
\hline
\texttt{Function} & \texttt{Moore} & \texttt{Filib} & \texttt{boost} & \texttt{P1788}\\
\hline
\texttt{sin}    &  0.552 &  0.032 &  1.444 &  9.320 \\
\texttt{cos}    &  0.156 &  0.032 &  1.560 & 10.172 \\
\texttt{tan}    &  0.124 &  0.020 &  0.756 &  2.476 \\
 \texttt{atan}  &  0.408 &  0.036 & 10.424 & 10.656 \\
 \texttt{exp}   &  0.308 &  0.164 &  4.532 &  4.644 \\
 \texttt{asin}  &  0.356 &  0.088 & 16.572 & 16.156 \\
 \texttt{acos}  &  0.368 &  0.088 & 16.724 & 16.748 \\
 \texttt{log}   &  0.272 &  0.044 &  5.404 &  5.272 \\
\hline
\end{tabular}
\end{table}

I emphasize that I tried to
be fair with all libraries and, to the
best of my knowledge, I used the faster options
for each library. For instance, I used the
boost library on its unprotected mode, which
does not change rounding modes in order to
evaluate arithmetical expressions. The code
was compiled with gcc 6.2.0 with flag
-O3 and \texttt{NDEBUG}  defined (the flag
-frounding-math should also be used when compiling
the Moore library.)

\section{Limitations}
The Moore library was designed and implemented using
a novel feature of the C++ language called
concepts \cite{cpts}, and it pays the
price for using the bleeding edge of this technology.
The main limitations in the library are due
to the current state of concepts in C++. For instance,
only the latest versions of the gcc compiler
support concepts, and today the
library cannot be used with other compilers.
Concepts are not formally part
of C++ yet, and it will take a few years
for them to reach their final form and become
part of the C++ standard.

Additionally, several decisions regarding
the library were made in order
to get around bugs in gcc's implementations of
concepts and in the supporting libraries,
and in order to reduce the compilation
time. The code would certainly be cleaner if
we did not care about these practical issues,
but without the compromises we took using the
library would be more painful.

Another limitation is the need to
guard the code by constructing an object of type \texttt{UpRounding}.
In other words, the code must look like this
{
\small
\begin{verbatim}
UpRounding r;
code using the Moore library
\end{verbatim}
}
\noindent
A similar requirement is made by the most efficient
rounding policy for the boost library, but that
library allows users to choose other policies for rounding,
although the resulting code is less efficient.
Things are different with the Moore library:
as the buyers of Henry Ford's cars in the 1920s, its
users can choose any rounding mode as they want,
so long as it is upwards. Users wanting
to mix code from the Moore library with
code requiring rounding to nearest will
need to resort to kludges like this one:
{\small
\begin{verbatim}
{
UpRounding r;
do some interval operations
}
back to rounding to nearest
{
UpRounding r;
do more interval operations
}
\end{verbatim}
}
\noindent

\bibliographystyle{amsplain}
\bibliography{moore_fortaleza}

\end{document}